\documentclass{article}
\usepackage{emulateapj,pstricks,apjfonts}

\righthead{SARAZIN, ET AL.}
\lefthead{X-RAY OBSERVATIONS OF GLOBULAR CLUSTER NGC~288}

\slugcomment{Astrophysical Journal, in press}
\submitted{Astrophysical Journal, in press}

\begin{document}

\title{ROSAT HRI X-ray Observations of the Open Globular Cluster NGC~288}

\author{Craig L. Sarazin\altaffilmark{1},
	Jimmy A. Irwin\altaffilmark{2,3},
	Robert T. Rood\altaffilmark{1},
	Francesco R. Ferraro\altaffilmark{4,5},
	and \\
	Barbara Paltrinieri\altaffilmark{6,4}}

\altaffiltext{1}{Department of Astronomy, University of Virginia, 
P.O. Box 3818, Charlottesville, VA 22903-0818;
cls7i@virginia.edu,
rtr@virginia.edu.}

\altaffiltext{2}{Department of Astronomy, University of Michigan,
Ann Arbor, MI 48109-1090;
jirwin@astro.lsa.umich.edu.}

\altaffiltext{3}{Chandra (AXAF) Fellow.}

\altaffiltext{4}{Osservatorio Astronomico di Bologna, via Ranzani 1, I-40126
Bologna, Italy;
ferraro@astbo3.bo.astro.it.}

\altaffiltext{5}{European Southern Observatory, Karl Schwarzschild Strasse 2,
D-85748 Garching bei M\"unchen, Germany.}

\altaffiltext{6}{Istituto di Astronomia -- Universit\'a La Sapienza,
via G. M. Lancisi  29, I-00161 Roma, Italy; barbara@coma.mporzio.astro.it.}

% \altaffiltext{7}{Stazione Astronomica di Cagliari, 09012 Capoterra, Italy.}

\begin{abstract}
A {\it ROSAT} HRI X-ray image was obtained of the open globular cluster
NGC~288, which is located near the South Galactic Pole.
This is the first deep X-ray image of this system.
We detect a Low Luminosity Globular Cluster X-ray source (LLGCX)
RXJ005245.0$-$263449 with an X-ray luminosity of
$( 5.5 \pm 1.4 ) \times 10^{32}$ ergs s$^{-1}$ (0.1--2.0 keV),
which is located very close to the cluster center.
There is evidence for X-ray variability on a time scale of $\la$1 day.
The presence of this LLGCX in such an open cluster suggests
that dense stellar systems with high interaction rates are not needed
to form LLGCXs.
We also searched for diffuse X-ray emission from NGC~288.
Upper limits on the X-ray luminosities are
$L_X^h < 9.5 \times 10^{32}$ ergs s$^{-1}$ (0.52--2.02 keV) and
$L_X^s < 9.3 \times 10^{32}$ ergs s$^{-1}$
(0.11--0.41 keV).
These imply upper limits to the diffuse X-ray to optical light ratios
in NGC~288 which are lower than the values observed for X-ray faint
early-type galaxies.
This indicates that the soft X-ray emission in these galaxies
is due either to a component which is not present in globular clusters
(e.g., interstellar gas, or a stellar component which is not found in low
metallicity Population II systems), or to a relatively small number of bright
Low Mass X-ray Binaries (LMXBs).
\end{abstract}


\section{Introduction} \label{sec:intro}

Globular clusters are thought to have two distinct populations of X-ray
sources (e.g., Hertz \& Grindlay 1983).
The high luminosity X-ray sources have
$L_X > 10^{34.5}$ ergs s$^{-1}$ and are thought to be binary systems with an
accreting neutron star (the so-called Low Mass X-ray Binaries or LMXBs;
e.g., Verbunt 1996).
%  because of their X-ray bursts.
The nature of Low Luminosity Globular Cluster X-ray sources
(LLGCXs) with $L_X < 10^{34.5}$ ergs s$^{-1}$ has been elusive.
While a neutron star
origin has not been ruled out for the more luminous of the LLGCXs, they
are more commonly thought to be associated with white dwarf systems,
perhaps related to cataclysmic variables (CVs) which might be produced by
stellar interactions in dense cluster cores
(Di Stefano \& Rappaport 1994).
Such a model is not without difficulties; e.g., clusters with high
interaction rates are rich in neither CVs nor LLGCXs
(Shara \& Drissen 1995;
Johnston \& Verbunt 1996). 
It would be useful to search for X-ray sources in relatively
open globular clusters where stellar interactions are less likely.

Part of the difficulty in determining the origin of LLGCXs arises because
of the problem of finding their optical counterparts in crowded globular
cluster fields.
If accurate X-ray positions are known and the optical field is not too crowded,
it may be possible to associate the X-ray sources with unusual optical or
UV objects.
For example, blue variable (flickering) objects (candidate CVs) have
been found to fall within the error boxes of some
LLGCXs
(Paresce, De Marchi, \& Ferraro 1992;
Cool et al.\ 1993, 1995).
Alternatively, unusual optical/UV colors may be used to identify the
LLGCX.
Recently, a number of stars with very unusual and strong
UV excesses have been found which appear to be associated with LLGCXs in the
globular clusters M13
(Ferraro et al.\ 1998a)
and M92
(Ferraro et al.\ 1998b).
These objects lie far from the traditional globular cluster
sequences in UV color-magnitude diagrams.
In M13, there are only 3 such objects in a sample of $>$12,000 stars.
Two of the three UV outliers are within a few arcsec of
the positions of X-ray sources observed about 40\arcsec\ from the center
of M13
(Fox et al.\ 1996).
The third object has no associated X-ray peak, but LLGCXs are
known to be highly variable
(Hertz, Grindlay, \& Bailyn 1993).
If this connection between UV objects and LLGCXs could be solidified, it
would produce a valuable new technique for studying these mysterious
objects.
To search for associations of LLGCXs and unusual optical/UV objects,
one requires accurately determined X-ray positions in regions of
moderate stellar density, and high resolution optical and
UV observations.
This suggests that one compare {\it ROSAT} HRI observations of relatively
open globular clusters with
{\it HST} photometry of the same regions.

In addition to adding to our understanding of LLGCXs,
X-ray observations of globular clusters might lead to a better
understanding of the X-ray emission from elliptical galaxies.
In X-ray bright elliptical galaxies (galaxies with a high ratio
of X-ray to optical luminosity $L_X/L_B$), the soft X-ray emission is
mainly due to thermal emission from diffuse hot gas, which is not
generally present in globulars.
However, in X-ray faint ellipticals (galaxies with a low ratio
of X-ray to optical luminosity $L_X/L_B$), much of the emission might be
due to stellar sources.
Globular clusters are the
best local analogs in which we can observe and possibly determine
nature of the X-ray sources in old stellar populations.

This is particularly interesting for the very soft X-ray
component observed from X-ray faint elliptical galaxies.
In early-type galaxies, this soft component has an X-ray to optical
luminosity ratio of $L_X / L_B \approx 10^{29.6}$ ergs s$^{-1}$
$L_\odot^{-1}$
(0.1--2.0 keV;
Kim, Fabbiano, \& Trinchieri 1992;
Irwin \& Sarazin 1998a,b),
and a spectrum with a temperature of about 0.2--0.3 keV
(Fabbiano, Kim, \& Trinchieri 1994; Kim et al.\ 1996).
The origin of this elliptical galaxy soft X-ray component is very
uncertain.
First, is the emission due to stars or to interstellar gas?
If it is stellar, what is its origin?
Stellar sources known to have soft X-ray spectra such as M-dwarfs,
RS CVn stars, and super soft sources seemed like promising explanations,
but none of these sources have the required X-ray characteristics to
fully account for the soft X-ray emission
(Pellegrini \& Fabbiano 1994;
Irwin \& Sarazin 1998b).
However, recent work has suggested that LMXBs are a viable candidate for the
soft X-ray emission
(Irwin \& Sarazin 1998a,b).
If the soft emission in ellipticals is interstellar, it should not
be present in globular clusters.
Alternatively, since globulars contain an old stellar population which
is similar to that in ellipticals (albeit with lower metallicity), one
might expect that globulars would contain any stellar X-ray emitting
component found in ellipticals, as long as the X-ray emission was
not very strongly dependent on metallicity
(however, see Irwin \& Bregman 1998).

For a globular cluster with an optical luminosity of the order of
$10^5 \, L_\odot$,
the expected soft X-ray luminosity would be
$\approx 10^{34.6}$ ergs s$^{-1}$.
High spatial resolution X-ray observations of globulars can be used
to separate brighter X-ray sources from any diffuse background.
We will use the present observations to detect or limit the total diffuse
soft emission from NGC~288.
Such observations are particularly important to test stellar models for 
soft X-ray emission from ellipticals in which a large number of
individually faint stars (e.g., M dwarfs) would produce the observed
X-rays.
For the purpose of studying the soft X-ray component from ellipticals,
the best globulars are those with low stellar densities (to avoid the
effects of interactions of the stellar population) and low absorbing
columns.

Here, we report on new X-ray observations of the globular cluster
NGC~288.
There have been no previous {\it Einstein} or pointed {\it ROSAT}
observations of this cluster.
The {\it ROSAT} All Sky Survey only had an exposure of 341 s at this
location, and only gave a weak upper limit on the total X-ray flux of
$< 1.3 \times 10^{-13}$ ergs cm$^{-2}$ s$^{-1}$ (0.5--2.5 keV;
Verbunt et al.\ 1995).
We have a series of deep {\it HST} observations of NGC~288
including UV images, which will be presented
in a future paper.
In many ways, NGC~288 is an ideal cluster for the identification of
soft X-ray sources and the comparison to optical/UV candidates.
First, it is a fairly open cluster, with a core radius of
$r_c = 85$\arcsec\ and a half-light radius of
$r_{1/2} = 135$\arcsec\ 
(Trager, Djorgovski, \& King 1993)
This makes it easier to identify X-ray sources.
It is also less likely that binary sources have been affected by core
collapse, and more likely that any sources are due to the original
binary population.
Although initially it was expected that LLGCXs would be associated with
clusters with large stellar interaction rates,
the trend in that direction is not particularly strong
(Johnston \& Verbunt 1996).
Finding LLGCXs in NGC~288 could be especially pertinent to understanding
the soft X-ray emission from elliptical galaxies.
Second, NGC~288 is near the South Galactic Pole ($b^{II} = - 89.3^\circ$),
which reduces the effects of interstellar absorption on the spectrum and
diminishes the chances of superposed Galactic X-ray sources.
The reddening of the cluster is fairly low ($E_{B-V} = 0.03$;
Peterson 1993), which corresponds to an absorbing column to the
cluster of about $N_H = 1.6 \times 10^{20}$ cm$^{-2}$,
using the relation $N_H = 5.3 \times 10^{21} E(B-V)$
(Predehl \& Schmitt 1995).
At a distance of $d = 8.4$ kpc toward the South Galactic Pole
(Peterson 1993),
the cluster is likely to lie behind essentially all of the Galactic gas.
The total Galactic H~I column in this direction is
$N_H = ( 1.5 \pm 0.1 ) \times 10^{20}$ cm$^{-2}$
(Stark et al.\ 1992),
which is consistent with the reddening of the cluster.
The mass and optical luminosity of the cluster are
$M = 8 \times 10^4$ $M_\odot$ and $L_V = 4.0 \times 10^4$ $L_\odot$,
respectively.

In \S~\ref{sec:data}, we discuss the X-ray observations.
The resulting X-ray sources are listed in \S~\ref{sec:srcs}.
We derive a limit on the diffuse X-ray flux from NGC~288
in \S~\ref{sec:diffuse}.
Finally, our conclusions are summarized in \S~\ref{sec:conclude}.

\section{X-ray Observation} \label{sec:data}

The globular cluster NGC~288 was observed with the {\it ROSAT}
High Resolution Imager (HRI) during the period 6-7 January, 1998.
The total exposure time was 19,891 s, which is reduced to 19,692 s
after correction for deadtime.
In addition to the normal processing of the data, we examined the
light curve of a large source-free region to check for any periods
of enhanced background, and none were found.  
Because we are interested in accurate positions for any X-ray sources
located within the globular cluster,
we also examined the aspect history for any anomalies during
the accepted time in the image.
None were found.
These data were taken during a period when the standard data pipeline
processing included an error in the boresight, but the data presented
here were reprocessed after this error was corrected.

% SUBMISSION
% \placefigure{fig:xray}
% PREPRINT
\centerline{\null}
\vskip2.8truein
\includegraphics{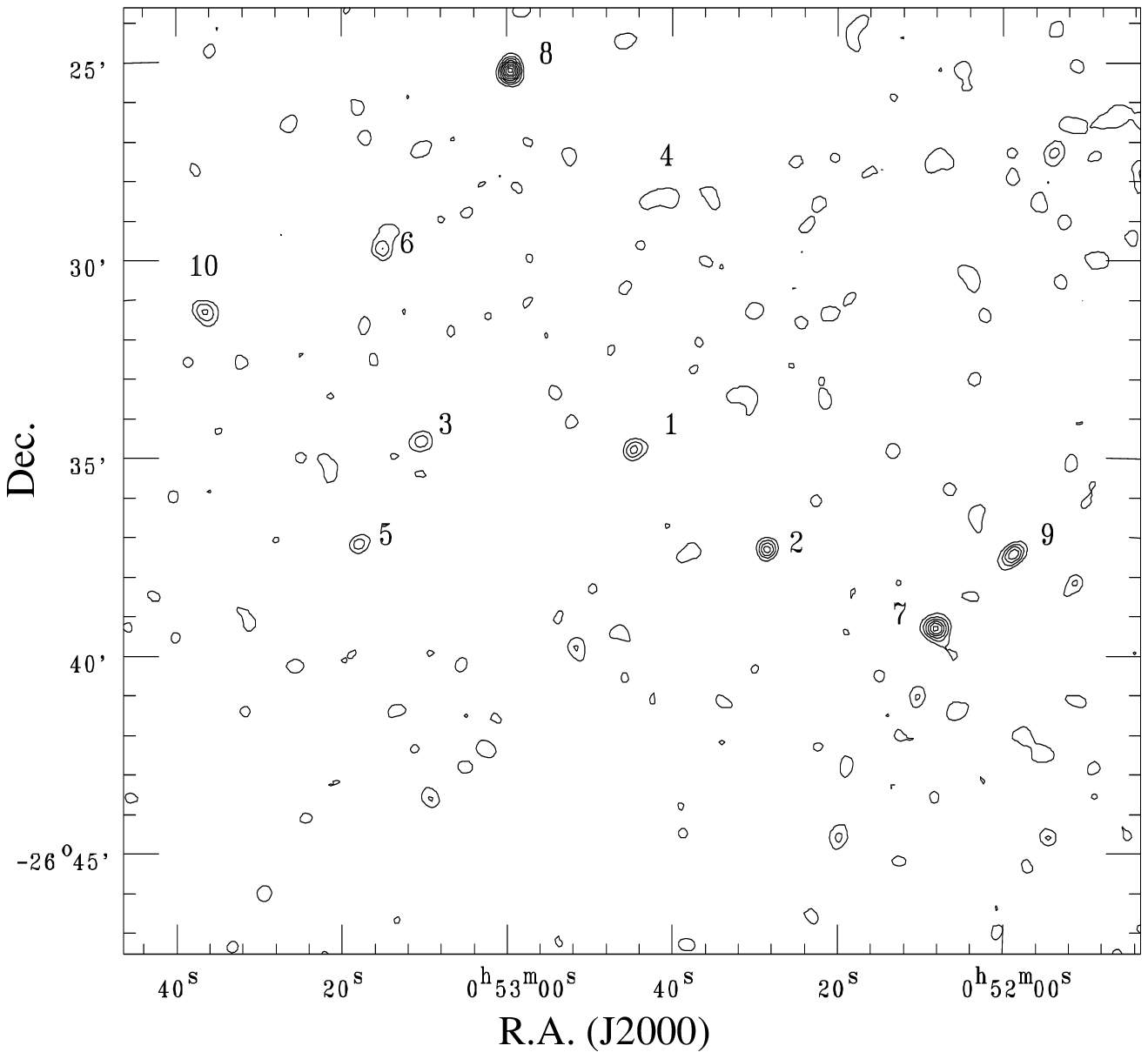}
\figcaption[fig1.ps]{
A contour plot of the {\it ROSAT} HRI image of the field around NGC~288.
There are nine contours ranging from 0.002 to 0.018 cnts s$^{-1}$
arcmin$^{-2}$.
The positions of the X-ray sources listed in
Table~\protect\ref{tab:srcs} are shown.
The center of the globular cluster NGC~288 is very near to the position
of Src.~1.
\label{fig:xray}}
% PREPRINT
\vskip0.2truein

The X-ray image was corrected for particle background, exposure, and
vignetting using the {\sc SXRB} software package of Snowden (1995).
For the purposes of display, the image was adaptively smoothed 
to a minimum signal-to-noise ratio of 5 per smoothing beam
(Huang \& Sarazin 1996).
Each smoothing beam was also required to have a larger FWHM than
the {\it ROSAT} HRI Point Spread Function (PSF).
A contour plot of the inner approximately 25\arcmin$\times$25\arcmin\ region
of the X-ray image is shown in Figure~\ref{fig:xray}.
Several point sources are obvious (\S~\ref{sec:srcs}), but
there is no evidence for any extended diffuse emission 
(\S~\ref{sec:diffuse}).
The center of the globular cluster NGC~288 is located very near to the
position of Src.~1 in this image.

\section{X-ray Sources} \label{sec:srcs}

Maximum-likelihood and local detection algorithms were used to
detect point sources in the HRI image.
A final detection criterion of 3$\sigma$ was adopted.
Table~\ref{tab:srcs} list the 10 sources which were detected.
The sources are also labeled on Figure~\ref{fig:xray}.
For each source, the Table gives its position,
its count rate and the 1$\sigma$ error,
its final signal-to-noise ratio SNR,
its projected distance $D$ from the center of the globular cluster,
and a comment on the identification.
The source count rates were corrected for the instrument PSF,
for background, and for vignetting.
The positions in Table~\ref{tab:srcs} are at the epoch J2000.
The statistical errors in the positions are generally less than 3\arcsec;
there may be a similar systematic error in the {\it ROSAT} HRI absolute
positions.
None of the sources were clearly extended;
however, for sources far from the center of the field where the
instrumental point-spread-function is very broad, the upper
limits on their size are large.
Src.~4 appeared possibly to be extended, but was also near the
detection limit.
Srcs.~8, 9, and 10 are identified with the quasars
QSO~0050.5-2641,
QSO~0049-2653,                
and
QSO~0051-267,
respectively.
The X-ray positions agree with the optical positions of these quasars
to better than 4\arcsec.
Src.~7 may be associated with the blue stellar object PHL~3043, which is
also likely to be quasar or other distant AGN.
We tried to use the positions of the known quasars to improve the
absolute positions of all of the sources, including the central X-ray
source located near the cluster center.
We determined more accurate optical positions for the quasars from
the Digital Sky Survey (DSS) image.
However, the differences between the optical and X-ray positions of these
three quasars do not show any simple systematic pattern, and are consistent
with the expected random measurement errors.
Thus, they do not allow us to improve the positions of the other sources.

% SUBMISSION
% \placetable{tab:srcs}
\begin{table*}[htb]
\caption{\hfil X-ray Sources \label{tab:srcs} \hfil}
\begin{center}
\begin{tabular}{lcccccl}
\tableline
\tableline
Src.&R.A.   &Dec.   &Count Rate&SNR&$D$&ID\\
    &(h:m:s)&(d:m:s)&(cts/ksec)&   &(arcmin)\\
\tableline
1 &00:52:45.0&$-$26:34:49&1.59$\pm$0.40&4.0&\phn0.12&RXJ005245.0$-$263449 in NGC~288\\
2 &00:52:28.6&$-$26:37:18&1.57$\pm$0.41&3.8&\phn4.52&\\
3 &00:53:10.4&$-$26:34:36&1.20$\pm$0.40&3.0&\phn5.64&\\
4 &00:52:41.0&$-$26:28:24&1.37$\pm$0.46&3.0&\phn6.39&\\
5 &00:53:18.1&$-$26:37:11&1.40$\pm$0.40&3.5&\phn7.76&\\
6 &00:53:15.3&$-$26:29:38&1.83$\pm$0.53&3.5&\phn8.44&\\
7 &00:52:08.4&$-$26:39:19&3.38$\pm$0.61&5.6&\phn9.42&PHL~3043 ?\\
8 &00:52:59.7&$-$26:25:10&4.85$\pm$0.74&6.6&   10.09&QSO~0050.5-2641\\
9 &00:51:58.8&$-$26:37:26&2.37$\pm$0.63&3.8&   10.72&QSO~0049-2653\\
10&00:53:36.4&$-$26:31:15&2.30$\pm$0.66&3.5&   11.97&QSO~0051-267\\
\tableline
\end{tabular}
\end{center}
\end{table*}

In Table~\ref{tab:srcs}, the sources are ordered by increasing distance
$D$ from the center of NGC~288.
We assume that the center is located at
R.A.\ = 00$^{\rm h}$52$^{\rm m}$45\fs3
and Dec.\ = $-$26\arcdeg34\arcmin43\arcsec\ 
(J2000;
Webbink 1985).
Note that this position differs considerably from the more recent
determination by
Shawl \& White (1986).
Comparison to the Digital Sky Survey (DSS) image of the cluster
shows that the Shawl \& White position does not agree with the apparent
center of the cluster, at least within the cluster core, while 
the Webbink position agrees reasonably well with the DSS image.

The count rate limit corresponding to our detection threshold is about
$1.2 \times 10^{-3}$ cts s$^{-1}$ near the center of the image;
vignetting and broadening of the PSF at large distances from the center
of the field increase the detection threshold there.
We have estimated the number of serendipitous X-ray sources expected
in this HRI observation using the deep source counts in
Hasinger et al.\ (1998).
For comparison, the count rate was converted into an unabsorbed
physical flux using the same assumptions as in Hasinger et al.\ (1998),
but assuming an absorbing column of $N_H = 1.6 \times 10^{20}$ cm$^{-2}$.
Based on the source counts in Hasinger et al., we would expect about
$8 \pm 4$ serendipitous sources in our observations.
This is obviously consistent with the observed number of 10 sources;
there is no overall excess in the number of X-ray sources in this
field.

% SUBMISSION
% \placefigure{fig:dss}
% PREPRINT
\begin{figure*}[htb]
\vskip4truein
\includegraphics{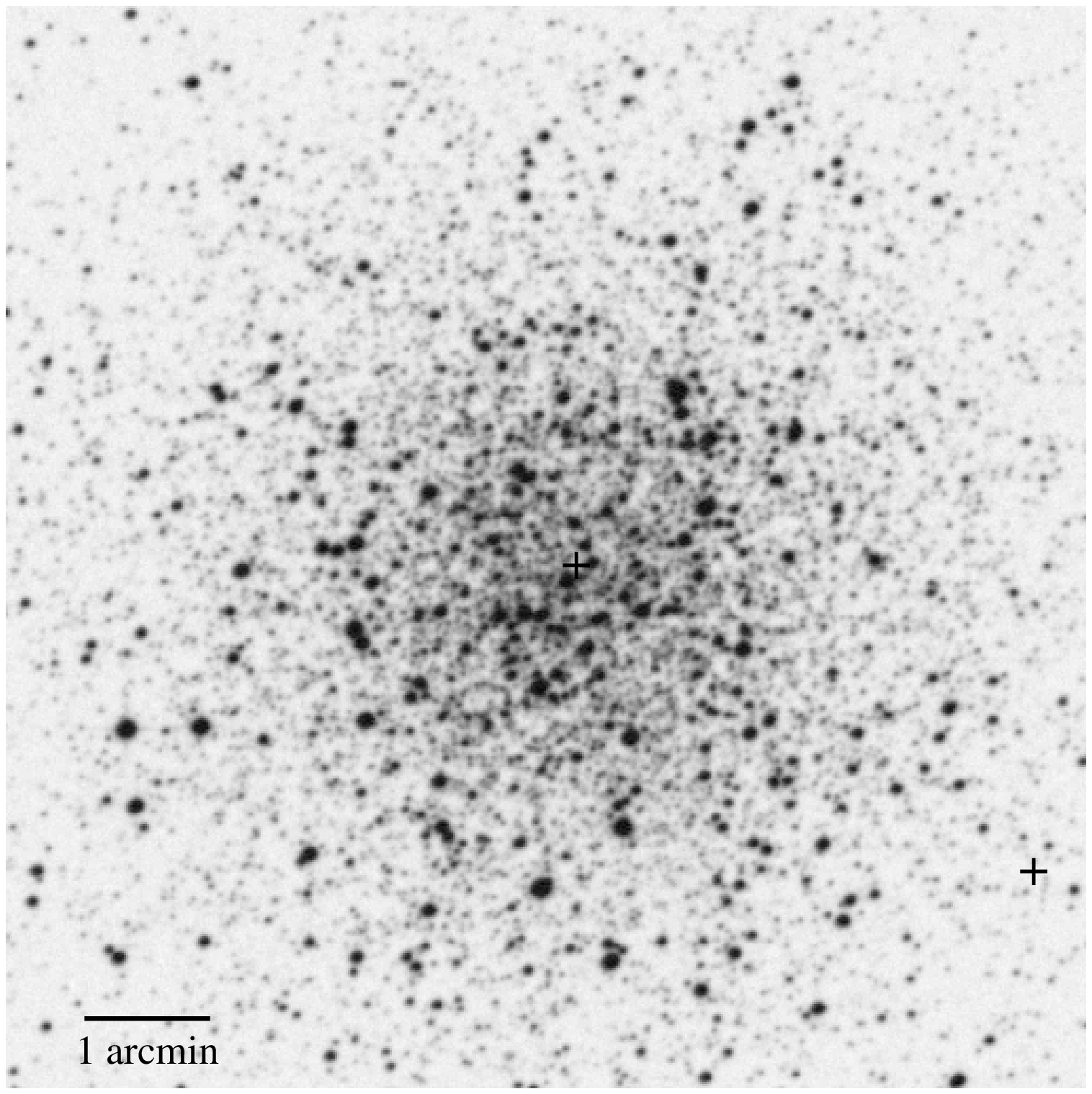}
\caption[fig2.ps]{
The Digital Sky Survey image of an approximately
9\arcmin$\times$9\arcmin\ field around the center of NGC~288.
The cross near the center of the cluster gives the position of Src.~1, which
is
RXJ005245.0$-$263449.
The cross at the lower right is the position of Src.~2.
The crosses are about twice the size of the error bars in the positions of
the sources.
\label{fig:dss}}
\end{figure*}

Figure~\ref{fig:dss} shows the optical image of NGC~288 from the DSS.
The small crosses near the center and at the lower right side of the
image show the positions of
Src.~1 (RXJ005245.0$-$263449) and Src.~2.
When examined at higher resolution, we find that the DSS image does
not show any very bright star which is within 5\arcsec\ of the position of
either Src.~1 or 2.

With the exception of Src.~1, the sources appear randomly distributed
throughout the HRI image, and are not concentrated to NGC~288.
On the other hand, Src.~1 (RXJ005245.0$-$263449) is located very close
to the center of the globular cluster.
It is unlikely that an unrelated foreground or background source would
be accidentally projected this close to the center of the cluster.
Including the slight decrease in the detector sensitivity to sources
at large radii from the center of the cluster, the probability that such
a close association would occur at random is less than 0.2\%.
If we increase the projected distance of RXJ005245.0$-$263449 from the
cluster center by the maximum amount permitted by the errors in the
X-ray and cluster center positions, the probability of a source
being this close to the center by accident is still less than
1\%.
Thus, we believe that RXJ005245.0$-$263449 is associated with the
globular cluster NGC~288.

On the other hand, the other sources (including Src.~2) are located at
projected distances from the cluster center at which serendipitous sources
would be likely.
All of these sources are at projected radii of at least two half-light
radii ($r_{1/2} = 135$\arcsec) where the density of cluster stars is
low.
Thus, although it is possible that Src.~2 (or some of the others) might
be associated with NGC~288, it seems much more likely that they
are only serendipitous foreground or background sources.

If we model the spectrum of 
RXJ005245.0$-$263449
as 1 keV thermal bremsstrahlung with an absorbing column of
$N_H = 1.6 \times 10^{20}$ cm$^{-2}$, the count rate of
$( 1.59 \pm 0.40 ) \times 10^{-3}$ s$^{-1}$ corresponds to an
unabsorbed flux of
$( 6.5 \pm 1.6 ) \times 10^{-14}$ ergs cm$^{-2}$ s$^{-1}$
(0.1--2.0 keV).
The flux is only weakly dependent on the form of the spectrum
assumed.
This implies a luminosity of $L_X = ( 5.5 \pm 1.4 ) \times 10^{32}$
ergs s$^{-1}$ (0.1--2.0 keV), which places
RXJ005245.0$-$263449
among the LLGCXs.

% SUBMISSION
% \placefigure{fig:time}
% PREPRINT
\begin{figure*}[htb]
\vskip2.0truein
\includegraphics{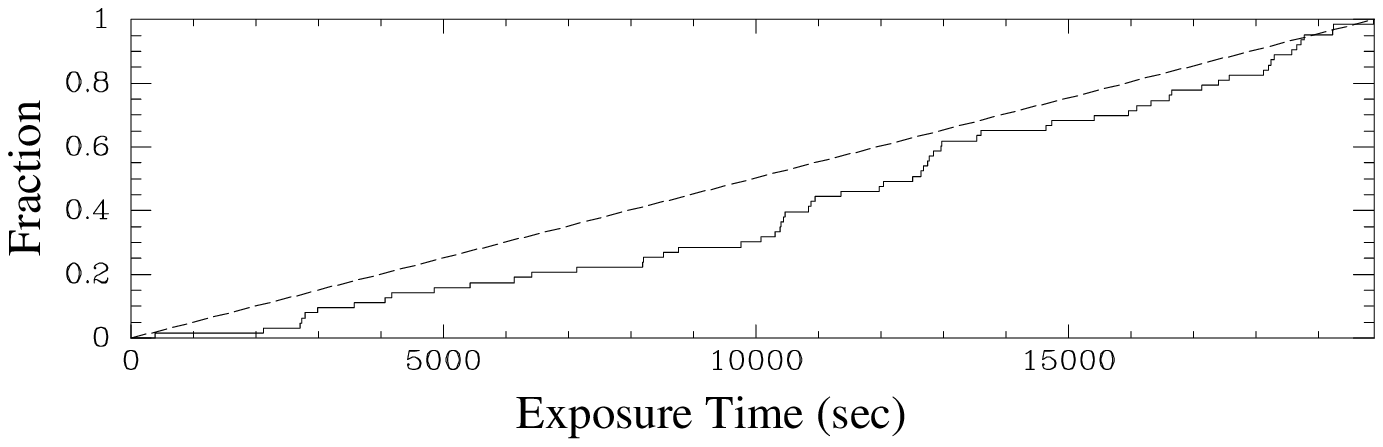}
\caption[fig3.ps]{
The arrival times of the photons from the X-ray source
RXJ005245.0$-$263449
near the center of NGC~288.
The abscissa gives the cumulative exposure time, while the ordinate gives
the cumulative fraction of the photons from the source and background
which have arrived by that time.
The solid histogram gives the data, while the dashed straight line is the
prediction if the source and background are constant.
\label{fig:time}}
\end{figure*}

RXJ005245.0$-$263449 appears to be a variable X-ray source.
Figure~\ref{fig:time} shows a histogram of the cumulative fraction of
the counts from the 20\arcsec\ radius circle centered on the source as
a function of the cumulative exposure time.
The dashed line shows the expected linear increase with exposure time,
assuming the source and background are constant.
The assumption that the source and background are constant can be rejected
at a confidence level of greater than 95\% for either the
Kolmogorov-Smirnov or Cramer-von Mises test.
Figure~\ref{fig:time} shows that most of the photons arrived during the
end of the cumulative exposure time.
Because the exposure time of 19,891 s was broken into 5 observing
intervals spread over about $1.2 \times 10^5$ s, this indicates that
the source in probably variable on a time scale of $\la$1 day.
Unfortunately, the number of source counts is too low to allow
any more detailed analysis of the source variability.

% PREPRINT
% ?? remove
\centerline{\null}
\vskip0.1truein

\section{Limit on Diffuse X-ray Emission} \label{sec:diffuse}

We also used the {\it ROSAT} HRI observations to place a limit on
any diffuse X-ray emission from the globular cluster.
We determined the emission from within the projected area of the
half-light radius of the cluster, which is $r_{1/2} = 135\arcsec$.
Any point sources were removed from within the regions used for determining
the cluster emission and the background.
The background was determined in two ways.
First, the Snowden (1995) {\sc SXRB} routines were used to determine the
particle background and the exposure map of the HRI image.
The particle background was removed from the image, and the image was
flat-fielded by dividing by the exposure map.
The X-ray background was determined from an annulus from 8\arcmin\ to
15\arcmin.
Second, the background was determined from the background image provided
with the standard data products, using the same region as was used to
collect the counts from the cluster.
These two methods gave consistent results, but the second technique gave
a more conservative (larger) upper limit on the X-ray flux, which is
the one we adapt.

No diffuse emission was detected from the cluster; the net count rate
was actually very slightly negative (although consistent with zero).
The 90\% confidence upper limit on the count rate from within the
half-light radius was $\le 4.5 \times 10^{-3}$ counts s$^{-1}$.
As discussed above, we adapt a hydrogen column to the cluster of
$N_H = 1.6 \times 10^{20}$ cm$^{-2}$.
In order to permit a more detailed comparison to the soft X-ray emission of
X-ray faint elliptical galaxies and to that expected from M stars (which
also have a very soft spectrum),
the diffuse emission is modeled as a 0.2 keV thermal spectrum with solar
abundance.
The upper limit on the count rate corresponds to an unabsorbed flux
limit of
$F_X^h < 1.13 \times 10^{-13}$ ergs cm$^{-2}$ s$^{-1}$
in the 0.52--2.02 keV {\it ROSAT} hard band, and
$F_X^s < 1.10 \times 10^{-13}$ ergs cm$^{-2}$ s$^{-1}$ in the
0.11--0.41 keV {\it ROSAT} soft band.
The hard band flux limit is nearly independent of the spectrum assumed,
but the soft band limit is affected by the spectral model.
At an assumed distance of $d = 8.4$ kpc, these flux limits correspond to
luminosity limits of
$L_X^h < 9.5 \times 10^{32}$ ergs s$^{-1}$ ({\it ROSAT} hard band
0.52--2.02 keV), and
$L_X^s < 9.3 \times 10^{32}$ ergs s$^{-1}$ ({\it ROSAT} soft band
0.11--0.41 keV).
If the luminosity of RXJ005245.0$-$263449 is added to these numbers to
give a limit on the total X-ray luminosity of the cluster (either diffuse
of resolved), the luminosity limits are increased by 35\%.
The total $B$-band optical luminosity of the cluster is
$L_B = 3.91 \times 10^4 \, L_\odot$
(Peterson 1993),
and the luminosity from within the projected half-light radius
is obviously one-half of this value.
This leads to  upper limits on the diffuse X-ray--to--optical luminosity
ratio of the cluster of
$L_X^h / L_B  < 4.9 \times 10^{28}$ ergs s$^{-1}$ $L_\odot^{-1}$ and
$L_X^s / L_B  < 4.8 \times 10^{28}$ ergs s$^{-1}$ $L_\odot^{-1}$.

\section{Conclusions} \label{sec:conclude}

We have obtained the first deep X-ray image of the open globular 
cluster NGC~288, which is located near the South Galactic Pole.
We detect a Low Luminosity Globular Cluster X-ray (LLGCX) source
RXJ005245.0$-$263449, which is located within $\sim$10\arcsec\ of
the cluster center.
The X-ray luminosity is $L_X = ( 5.5 \pm 1.4 ) \times 10^{32}$
ergs s$^{-1}$ (0.1--2.0 keV).
There is evidence that RXJ005245.0$-$263449 is variable, and that the
X-ray flux increased during the $\sim$1 day period of the observation.
The fact that this LLGCX is present in such an open cluster adds evidence
to the argument that dense stellar systems with high interaction rates
are not needed to form LLGCXs
(Johnston \& Verbunt 1996). 
On the other hand, the fact that RXJ005245.0$-$263449 is located so
close to the cluster center (its projected distance from the center is
less that one-tenth of the core radius) is consistent with its being
a binary system more massive than the typical cluster star.

A series of deep {\it HST} images of NGC~288 have been obtained
(observation 6804) and are now being analyzed;
these include UV images.
It may be possible to identify RXJ005245.0$-$263449 on these images.
It should also be possible to test the hypothesis that LLGCXs are
associated with stars with extremely blue UV colors
(Ferraro et al.\ 1998a).
Obviously, one would expect RXJ005245.0$-$263449 to be identified with
such a UV bright star.
Also, one would not expect to find a large number of other UV bright
sources in the {\it HST} image, since RXJ005245.0$-$263449 is the
only X-ray source detected in the central region of the cluster.
On the other hand, LLGCXs are highly variable, so the detection of
a single UV star without an X-ray counterpart would not prove that
LLGCXs are not associated with UV stars.
For example, RXJ005245.0$-$263449 seems to be variable, and might not have
been detected if it had been faint throughout our observation.

We also searched for diffuse X-ray emission from NGC~288.
Upper limits (90\%) on the X-ray luminosities of
$L_X^h < 9.5 \times 10^{32}$ ergs s$^{-1}$ for the {\it ROSAT} hard band 
(0.52--2.02 keV) and
$L_X^s < 9.3 \times 10^{32}$ ergs s$^{-1}$ for the {\it ROSAT} soft band 
(0.11--0.41 keV)
are obtained within the optical half-light radius of the cluster,
$r_{1/2} = 135\arcsec$.
These imply upper limits to the diffuse X-ray to optical light ratios
of
$L_X^h / L_B  < 4.9 \times 10^{28}$ ergs s$^{-1}$ $L_\odot^{-1}$ and
$L_X^s / L_B  < 4.8 \times 10^{28}$ ergs s$^{-1}$ $L_\odot^{-1}$.
These upper limits are lower than the values observed for most
X-ray faint early-type galaxies
(Irwin \& Sarazin 1998a,b).
This indicates that the soft X-ray emission in X-ray faint elliptical
galaxies is due to a component which is not present in globular clusters
(e.g., interstellar gas, or a stellar component which is not found in low
metallicity Population II systems), or that the soft emission comes from
a relatively small number of bright X-ray sources (e.g., LMXBs;
Irwin \& Sarazin 1998a).
If the emission were due to a small number of bright X-ray sources, then
the expected number in a typical globular would be less than unity.

\acknowledgments

We thank Flavio Fusi Pecci for helpful comments.
We would also like to the thank the referee for useful suggestions which
improved the presentation.
C. L. S. was supported in part by NASA grants NAG 5-4516, NAG 5-3057,
and NAG 5-8390.
R. T. R. is supported in part by
NASA Long Term Space Astrophysics Grant NAG 5-6403 and STScI/NASA
Grant GO-6607.
F. R. F. acknowledges the ESO Visiting Program for its hospitality.
The optical image of NGC~288 is from the Digital Sky Survey, which was
produced at the Space Telescope Science Institute using photographic data
obtained using the Oschin Schmidt Telescope on Palomar Mountain and
the UK Schmidt Telescope.

% SUBMISSION
% \clearpage

\end{document}